\documentclass[prl,floatfix,twocolumn]{revtex4}
\usepackage{amsmath}
\usepackage{amsfonts}
\usepackage{mathrsfs}
\usepackage{epsfig}
\usepackage{graphicx}
\usepackage{dcolumn}
\usepackage{hyperref}
\usepackage[hyphenbreaks]{breakurl}

\begin{document}
\title{Emergence of frustration signals systemic risk}
\author{Chandrashekar Kuyyamudi$^{1,2}$, Anindya S. Chakrabarti$^3$ and
Sitabhra Sinha$^{1,2}$}
\affiliation{$^1$The Institute of Mathematical Sciences, CIT Campus,
Taramani, Chennai 600113, India.\\
$^2$Homi Bhabha National Institute, Anushaktinagar, Mumbai 400094,
India.\\
$^3$Economics area, Indian Institute of Management, Vastrapur,
Ahmedabad 380015, India.
}
\date{\today}
\begin{abstract}
We show that the emergence of systemic risk in complex systems
can be understood from the evolution of functional networks
representing 
interactions inferred from fluctuation correlations
between macroscopic observables.
Specifically, we analyze the
long-term collective dynamics of the New York Stock Exchange between
1926-2016, showing that
periods marked by systemic crisis, viz., around the Great Depression of
1929-33 and the Great Recession of 2007-09, are
associated with emergence of frustration indicated by the loss of
structural balance in the interaction networks.
During these periods the dominant eigenmodes characterizing the
collective behavior exhibit delocalization leading to 
increased coherence in the dynamics.
The topological structure of the networks exhibits a
slowly evolving trend marked by the emergence of a prominent core-periphery
organization around both of the crisis periods.
\end{abstract}

\maketitle
Analyzing the collective behavior of a complex system comprising many
components is challenging as the interactions between the local
dynamics of the individual components often result in the emergence of
qualitatively different phenomena at the
systems-level~\cite{Hidalgo2009,May2013}. 
Indeed many natural and social
systems have extremely large number of constituent elements, each
following complicated trajectories and influencing others through
heterogeneous forms of coupling which may also evolve over
time~\cite{Castellano09,Sugihara2012}. Thus,
a complete microscopic description of such systems in terms of all the
variables involved will be intractable. Instead, a parsimonious
description in terms of effective interactions between relatively
fewer number of key macroscopic observables may prove to be more
effective in understanding the global evolution of the
system~\cite{Beale2011,Battiston2016}. The
nature of connectivity between the observables can be inferred from
the cross-correlations between their temporal evolution~\cite{Gomez2009}, 
allowing the
construction of a functional network coordinating the overall
behavior. This approach has been successfully used for analyzing many
natural (e.g., the brain~\cite{Chialvo2010,Arenas2016}) 
and socio-economic (e.g., financial
markets~\cite{Laloux1999,Plerou1999,Pan2007}) systems, 
where it can help us understand how extreme events, such as
large-scale failures leading to systemic crisis, can
arise~\cite{May2008,Haldane2011,Noel2013}.

Financial markets provide a particularly appropriate example where
such an approach can be used to identify possible universal patterns characterizing 
self-organizing collective phenomena 
because of the availability of a large volume 
of high-quality data~\cite{Mantegna1999,Bouchaud2003,Sinha11}. 
The global dynamics of such a system emerges from 
millions of transactions daily at the micro-scale between market participants, ranging from individual traders to institutional 
investors who are variously driven by changes in economic
fundamentals, herding effects, idiosyncratic motives and exogenous
shocks like news~\cite{Cont2000,Gabaix2006}.
A macroscopic description can thus be framed in terms of effective interactions between financial assets 
inferred from correlations in their price fluctuations that result from the aggregation of all transactions in these assets by market participants.
Here we have analyzed the collective dynamics of price fluctuations in the largest stock market, viz., New York Stock Exchange (NYSE), over  
a period of nine decades (1926-2016) to understand the evolution of
this complex system over extremely long
time-scales~\cite{Kuyyamudi2015}.
Specifically, we aim to identify quantitative signatures of systemic
risk whereby initially localized perturbations
can eventually trigger economy-wide
catastrophic changes~\cite{Acemoglu2015,Bardoscia2017}.

\begin{figure}[tbp]
\begin{center}
\includegraphics[width=0.99\linewidth,clip]{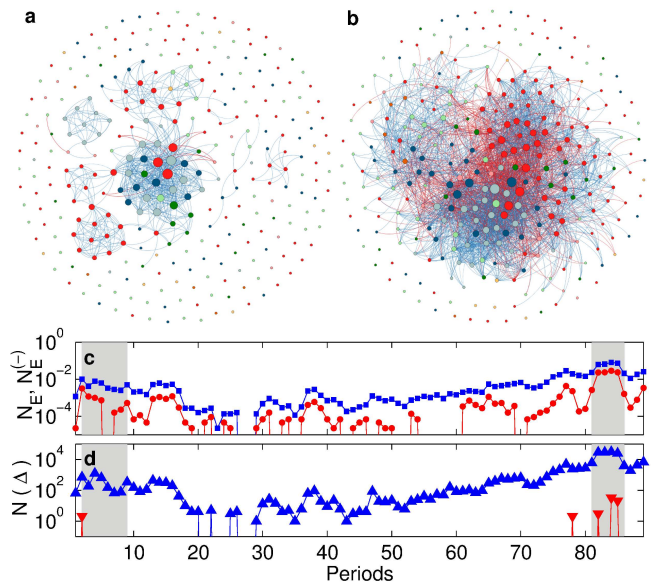}
\end{center}
\caption{(color online). 
(a-b) Interaction networks of 
stocks traded in the New York Stock Exchange (NYSE)
showing the top 300 in terms of average price
during (a) Period 80: Dec 2002-Dec 2006
and (b) Period 85: Feb 2008-Feb 2012, corresponding to time-intervals prior
to and during the
Great Recession. Edges represent significant interactions obtained
by spectral filtering of the cross-correlation matrix removing the
common ``global'' mode
and noise which mask these interactions. The crisis period
is characterized by the emergence of a large number of
negative edges (indicated by red color) representing anti-correlation between
pairs of stocks. This period also exhibits dense clustering evident from the
presence of many connected triads.
(c-d) The time-evolution of the number of total edges ($N_E$, in blue) and
negative edges ($N^{(-)}_{E}$, in red) is shown in (c) and that of the number
of connected triads [$N(\Delta)$, blue: all triads, red: frustrated
triads] is shown in (d) for 
the interaction networks of NYSE stocks between 1926-2016. 
The intervals corresponding to
systemic crisis (viz., the Great Depression of 1929-1933 and the Great
Recession of 2007-9) in the economy are indicated by the shaded
regions. The red inverted triangles indicate the periods where
structural balance in the interaction network
is lost.
}
\label{fig1}
\end{figure}

In this paper we uncover several important features of the long-term dynamics of the market by investigating the topological features characterizing the interaction networks between stocks over the entire sequence of different intervals comprising the period under study [see Fig.~\ref{fig1}(a-b) for
two representative networks from periods prior to and during the 2007-9 crisis].  
The most striking of these is the observation that periods of major economic crisis, viz.,
the Great Depression of 1929-33 and the Great Recession of 2007-9, are associated
with the loss of structural balance in the corresponding interaction network for those periods.
It is significant that such a phenomenon, which is equivalent to the emergence of frustration, is
often identified with a major regime transition in complex systems, e.g., in physical systems
(spin glass~\cite{Fischer1991,Stein2013}) as well as in social systems (such as the loss of structural balance in the
network of strategic alliances between nations preceding the outbreak
of the first world war~\cite{Antal2005}).
We show the existence of a slowly evolving trend in the properties of the
interaction network, most notably in the number of edges with negative weights which is related to
the intensity of anti-correlated movements [Fig.~\ref{fig1}(c)] and in the number of connected triads
that indicates the extent of clustering [Fig.~\ref{fig1}(d)]. These measures have low values for most of the duration under study but become
high close to the two periods of crisis mentioned earlier.
An associated result is that the networks corresponding to the crisis
periods exhibit stronger assortativity indicative of a prominent
core-periphery organization. We indeed observe large, densely
connected inner cores in the periods around the two crises, suggesting an
increased degree of coherence in the movement of stocks. This is
reinforced by our demonstration of delocalization in the strongest
eigenmodes characterizing large-scale correlated movements during
these times~\cite{note4}.

\begin{figure}[t!]
\begin{center}
\includegraphics[width=0.99\linewidth,clip]{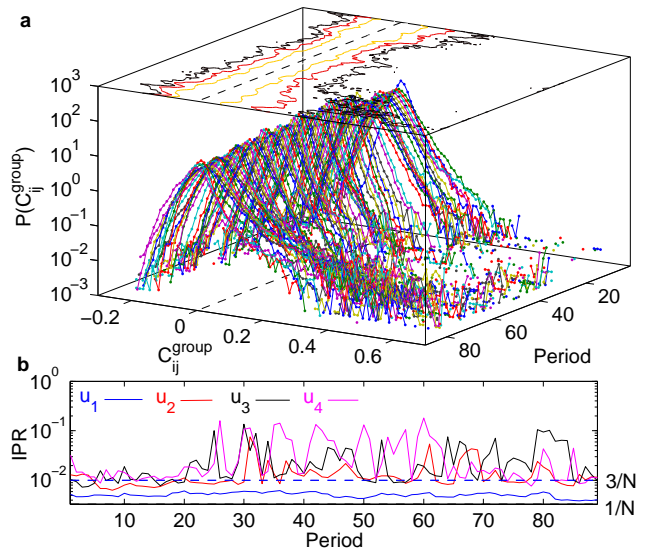}
\end{center}
\caption{(color online). 
(a) The evolution of the probability distribution of significant
correlations between stocks (after spectral filtering) during
1926-2016. The adjoining contour plot shows the temporal variation
in the dispersion of the distribution across the period of about 90
years, the different contours corresponding to integral multiples of
the standard deviation.
(b) The inverse participation ratio (IPR) of the global mode corresponding to the largest
eigenvalue ($u_1$) and that for the next few leading ``group'' modes corresponding to the second, third and fourth
highest eigenvalues ($u_2, u_3, u_4$). The lower bound of IPR,
viz.,
$1/N$ corresponds to a completely delocalized
mode in which all components contribute equally while the broken line
represents the expected IPR for random noise ($= 3/N$). The group
modes exhibit stronger localization during periods of relative
economic calm, while systemic crisis is associated with increased coherent movement among
stocks implied by delocalization. 
}
\label{fig2}
\end{figure}

The data used for our analysis is obtained from a database comprising time series of 
daily closing prices for all stocks traded in NYSE between December
31, 1925 and March 18, 2016, which is maintained by the Center for Research in Security Prices (CRSP)~\cite{CRSP}.
We have segregated the total duration into 89 overlapping periods of $T=1001$ days, with an overlap of 260 days (approximately corresponding to the number of working days in a year). 
In each of these periods we focus on $N=300$ stocks having the highest
average price in that interval among all the stocks~\cite{note1}. We
have repeated the analysis with different samples obtained from the
database, e.g., by increasing $N$ to 500 or choosing 300 stocks by
random sampling, and have obtained qualitatively
similar results.
To construct the interaction network between stocks for each period, we focus on cross-correlations
$C_{ij} = \langle r_i r_j \rangle$ ($i,j = 1, \ldots, N$) in the 
normalized logarithmic returns $r_{i,t}$ ($=[R_{i,t}-\langle R_i \rangle]/\sigma_i$, $\langle . \rangle$ and $\sigma$ representing mean and standard deviation, respectively) that measures
daily fluctuations in the stock prices $p_i$ ($R_{i,t}=\log(p_{i,t+1}/p_{i,t})$, $i=1\ldots, N$).
Spectral decomposition of the cross-correlation matrix {\bf C} allows it to be expressed as the
sum of $N$ terms (corresponding to the different eigenmodes), viz., $\sum_{\alpha} \lambda_{\alpha} u_{\alpha} u_{\alpha}^T$ ($\alpha = 1, \ldots, N$) where
$\lambda$ and $u$ are the eigenvalues and eigenvectors of {\bf C}. 
By comparing with the spectrum of the corresponding Wishart matrix, i.e., the random
correlation matrix constructed from $N$ mutually uncorrelated time-series each of length $T$,
we filter out from {\bf C} all modes corresponding to eigenvalues that lie within
the bounds $\lambda^W_{max,min}=[1 \pm (1/\sqrt{T/N)})]^2$ of the
Wishart matrix spectrum~\cite{Marchenko67}. 
These ``random'' modes are essentially indistinguishable from noise and we thus focus only on
the deviating modes whose eigenvalues are higher than $\lambda^W_{max}$. The eigenmode corresponding
to the largest eigenvalue is identified as the ``global'' mode that represents signals common to the
entire market, while the other deviating eigenmodes together represent the ``group'' dynamics 
between related stocks which are responding
to similar signals specific to the community to which they belong, e.g., stocks
in the same industrial sector. 
Thus, the spectral decomposition of the collective dynamics allows it to be 
split into three components, viz., {\bf C} = {\bf C}$^{global}$ + {\bf
C}$^{group}$ + {\bf C}$^{random}$. Fig.~\ref{fig2}~(a) shows the
time-evolution of the distribution of the elements of {\bf C}$^{group}$ over
the period under study.

The distinct nature of the eigenmodes belonging to these three
components is indicated by the qualitatively different distributions
followed by the corresponding eigenvector components. In particular,
one can quantify this difference by measuring the degree of
localization of the eigenmodes using their inverse participation ratio
(IPR). This is defined for the $k$-th
eigenvector $u_k$ as
$I_k = \sum_{i=1}^{N} [u_{ki}]^4$
where $u_{ki}$ is the $i$-th component of the vector~\cite{note3}.
A completely delocalized eigenmode where all components have exactly
identical contribution, viz., $u_{ki} = 1/\sqrt{N}$, will yield an
IPR value of $1/N$. We note that this is approximately the case for
the global eigenmode corresponding to the largest eigenvalue of {\bf
C} [Fig.~\ref{fig2}~(b)]. The ``random'' eigenmodes also do not
exhibit substantial localization having IPR values around $3/N$
[indicated by the broken line in Fig.~\ref{fig2}~(b)]. In contrast,
the ``group'' eigenmodes can show a high degree of localization, with
the highest possible value of IPR, viz., $I_k = 1$ corresponding to
a single component (say $i$) dominating the eigenmode, i.e., $u_{ki} = 1$ and
$u_{kj}=0$ for $j \neq i$. We note from Fig.~\ref{fig2}~(b) that for
most of the periods investigated, the IPR for the eigenmodes $u_2,u_3,u_4$
corresponding to the three highest deviating eigenvalues are indeed much
larger than those for the random and global modes. They signify the
existence of groups of stocks (viz., the ones having dominant
contribution to a deviating eigenmode) such that members of a group
have their dynamics more in synchrony with each other as compared to
the rest.
However, during the
two intervals in which there was systemic crises (viz., the Great
Depression and the Great Recession), the IPR for all three deviating
eigenmodes shown here become almost indistinguishable from that of the random
modes. This
substantial decrease of the IPR indicates the occurrence of
delocalization in the group eigenmodes during times of crises, which
in turn signifies that stocks lose any distinct dynamical identity.
Thus, systemic crises are marked by globally synchronized
collective dynamics characterized by delocalization in all eigenmodes.

In order to analyze this evolution in the nature of interactions
between stocks in greater detail, we have reconstructed signed,
undirected networks representing significant correlations between
stocks for each of the 89 periods under investigation [two instances
are shown in Fig.~\ref{fig1}~(a-b)]. 
To obtain the
adjacency matrix {\bf A} for such a network from the 
corresponding group correlation matrix {\bf C}$^{group}$, we impose a
threshold such that $A_{ij} = 1$ if $C_{ij}^{group} > \mu_{random} + 3
\sigma_{random}$, $A_{ij} = -1$ if $C_{ij}^{group} < \mu_{random} - 3
\sigma_{random}$ and $A_{ij} = 0$ otherwise. Here, $\mu_{random}
(\approx 0)$ and $\sigma_{random}$ are the mean and standard deviation
(respectively) of the distribution of $C^{random}$ matrix elements~\cite{note2}.
%
Our analysis of the networks for the 89 successive periods reveal
significant temporal variation in the frequency of edges with negative
weights between various stocks [Fig.~\ref{fig1}~(c)]. The number of
connected triads (linked by statistically significant
cross-correlations) that are observed also vary over a large range
spanning 5 orders of magnitude but become particularly large only
during the period preceding (and following) the two particularly
catastrophic events that occurred during the period under study, viz.,
the Great Depression of 1929 and the Great Recession of 2007
[Fig.~\ref{fig1}~(d)].
Most strikingly, the number of {\em frustrated} triads which
measures the extent of the loss of {\em structural balance} in the network
of interactions only show non-zero values in the periods just prior to
and during the two crisis periods [Fig.~\ref{fig1}~(d)].

The concept of structurally balanced networks was originally
introduced in the context of social interactions~\cite{Heider1946} and refers to
systems having positive or negative links arranged such that they do
not give rise to inconsistent relations within cycles in the network.
An example of arranging such links so that they give rise to an
inconsistency (resulting in frustration or equivalently, loss of balance)
occurs when three nodes $A$, $B$ and $C$ are connected to each other such that
the links between $A, B$ and $B, C$ are positive, but that between $A, C$ is
negative. If the node states can be in one of two states (e.g., ``up''
or ``down''), it is easy to see that no possible assignment of states
exist that satisfy all the given relations between the nodes. In
general, a system loses balance if triads of connected nodes possess
an odd number of negative relations. It is also known that a balanced
network can be always mapped to a system comprising two subnetworks,
with only positive interactions existing within each subnetwork, while
links between the two are exclusively negative~\cite{Cartwright1956}.

\begin{figure}
\begin{center}
\includegraphics[width=0.99\linewidth,clip]{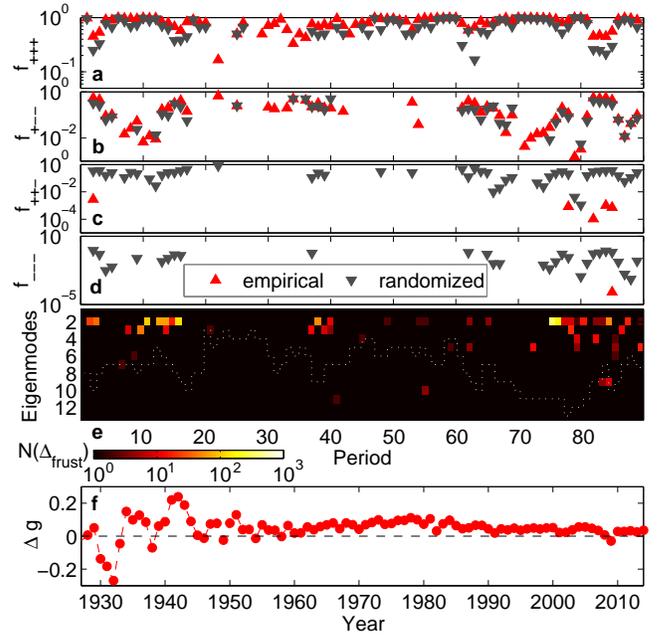}
\end{center}
\caption{(color online). 
(a-d) The evolution of the fraction of connected triads (red triangles) in the network of significant interactions resolved into the four possible distinct signed triads, (a) and (b) corresponding to balanced
(viz., $+++$ and $+--$, respectively), and (c) and (d) to unbalanced triads (viz., $++-$ and $---$).
For each, the fraction of triads expected from degree-preserved randomized networks are also 
shown (gray inverted triangles). Note that the unbalanced triads are much less likely to occur in the empirical
network.
(e) The number of frustrated triads [$N(\Delta_{frust})$, indicated by
the colorbar] in each of
the eigenmodes corresponding to the 13 largest eigenvalues of the
cross-correlation matrix for the periods 1-89 spanning the entire
duration being investigated. The white broken line indicates the
number of deviating eigenmodes $i$ (i.e., $\lambda_i > \lambda^{W}_{max}$)
in each period. 
While frustrated triads appear in individual eigenmodes in many
periods, the network of significant interactions exhibit such triads
only prior to and during systemic crisis in the economy [indicated by
negative growth rate, i.e., $\Delta g<0$, in (f)].
(f) Relative change $\Delta g$ in the gross domestic product (GDP) per
capita of USA measured in terms of
logarithmic returns over successive intervals for the same period as in (a-d). 
}
\label{fig3}
\end{figure}

To explore how the balance of the interaction network evolves over
different periods, we show in Fig.~\ref{fig3} the sign composition of
all the triads that occur in the network in a given period. The first
two panels [(a) and (b)] correspond to the fraction of balanced triads (i.e., those
having two or no negative edges) while the next two panels [(c) and
(d)]
show the fraction of unbalanced triads (i.e., having one or three
negative edges). We also show for comparison the fraction of triads of
each type that are expected to arise by chance given the degree
sequence of the network (calculated from the corresponding
signed degree-preserved randomized networks). 
We observe that in the empirical networks, unbalanced triads are far
less likely to occur than by chance (with the triad having all
negative links occurring
only once over the entire duration investigated), and are seen
exclusively during systemic crises in the financial system.
In Fig.~\ref{fig3}~(e) we show how each individual eigenmode $\alpha$
belonging to {\bf C}$^{group}$ (used for reconstructing the interaction network)
contributes to the loss of structural balance. 
This is done by computing the number of frustrated
triads in the network corresponding to the eigenmode $\alpha$. This
network is constructed from the matrix obtained through
the outer product $\lambda_{\alpha} u_{\alpha}^{T} u_{\alpha}$. We
observe that frustrated triads appear in individual eigenmodes in many
periods clustered around intervals having negative growth rate
measured in terms of relative change $\Delta g$ in the GDP per capita
that is shown in Fig.~\ref{fig3}~(f). It is worth noting that the
frustrated triads in the individual eigenmodes do not appear when we
consider the interaction network obtained by aggregating over these
modes except in the two major financial crises that occurred during
the entire duration under study.

\begin{figure}
\begin{center}
\includegraphics[width=0.99\linewidth,clip]{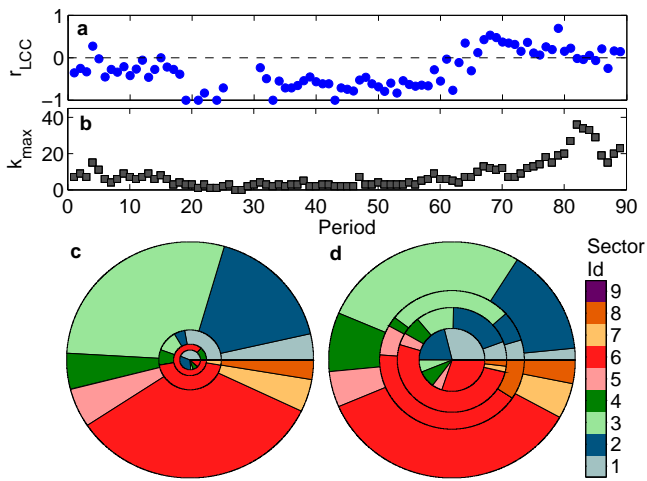}
\end{center}
\caption{(color online). 
(a) The evolution of the degree assortativity $r_{LCC}$ of the largest connected
component of the interaction networks 
over the duration under study. Periods associated with financial
crisis are seen to have
positive values of $r_{LCC}$ suggesting that the corresponding networks
can exhibit core-periphery organization. This is supported by
$k$-core analysis of the networks with panel
(b) showing the evolution of the order of the innermost core $k_{max}$ 
of the networks.
It is apparent that the depth of the core increases during systemic crisis, indicating
higher degree of coherence in the movement of stocks in the market. 
(c-d) Multi-level pie chart for
(c) Period 80 and (d) Period 85
showing the sector composition of each of the
$k$-shell quartiles.
The radii of the quartiles are proportional to
the fraction of stocks belonging to them. 
The color code in the adjoining key indicating the business sectors 
to which stocks belong correspond to
(1) Mining \& Construction, (2) Manufacture (basic), (3) Manufacture (advanced),
(4) Transportation, (5) Trade, (6) Finance, Insurance \& Real Estate,
(7) Services (Business), (8) Services (Public) and (9) Public
Administration.
}
\label{fig4}
\end{figure}

Analysis of the interaction networks (reconstructed from the group
correlation matrix) across time also indicates
changes in the macroscopic properties of the network 
which correlate with the stress in the economy. 
For instance, we observe that the network gets more connected during
crisis periods as indicated by the time-evolution of the number of
edges $N_{E}$ [Fig.~\ref{fig1}~(c)]. 
More intriguingly, there is evidence of {\em degree homophily}, i.e., nodes
tend to preferentially connect with other nodes having similar degree
(i.e., number of connections). This is measured for the largest
connected component of each of the networks using the measure of assortativity
coefficient $r_{LCC}$~\cite{Newman2003} [Fig.~\ref{fig4}~(a)]. It
allows us to observe a systematic variation in the mesoscopic
structure of the interaction network, with $r$ being positive (i.e.,
the network is assortative) during periods of stress and negative
(i.e., disassortative) in
other times. We note that degree assortativity in a network implies the existence
of a densely connected core of high degree nodes with low degree
nodes forming the periphery~\cite{Newman2003b}.

To characterize the core-periphery structure of the networks in detail
we have performed $k$-core decomposition~\cite{Seidman83} on them. This
technique successively reveals cores of higher order by sequentially
removing shells comprising nodes of lower degree 
in a recursive
manner. The order $k_{max}$ of the innermost core provides a measure of the
depth of interconnectedness in the system for a particular period.
Fig.~\ref{fig4}~(b) shows that $k_{max}$ peaks during the periods of
systemic crisis further underlining the earlier observation that
these periods are associated with global synchronization in the
collective dynamics. 
Fig.~\ref{fig4}~(c-d) shows the sector compositions of the shells by
grouping them into quartiles, for two of the interaction networks
corresponding to Periods 80 (pre-crisis) and 85 (crisis) respectively.
The fractional occupancy of a shell quartile by stocks belonging to
different industrial sectors is shown using a color code constructed
from the four digit Standard Industrial Classification (SIC) as has
been indicated in
the key.
In the pre-crisis period most of the stocks occupy the outer
periphery of the network, while during crisis the inner cores become
densely populated as indicated by the increased radius of the
innermost shell quartile. 
The diverse sectoral composition of the inner cores
is consistent with the emergence of system-wide coherent
activity during the crises periods as was suggested by the
delocalization of group eigenmodes. This is a quantitative
demonstration of the fact that the 2007-09 crisis spilled over to the
entire economy after originating in the financial
sector~\cite{Acharya2009}.

It is tempting to conclude that the loss of structural balance in the
interaction networks characterizing a financial market
can act as a robust indicator signaling the onset of a systemic crises. 
As already noted earlier, we have obtained qualitatively similar
results for other representative samples obtained from the same
database.
While it may appear that the loss of balance occurs as a sudden event
that coincides with the onset of systemic crisis in the system, the
other results reported here show that it is in fact connected to a
gradual build-up of stress in the market. The resulting change in the
nature of the collective dynamics is reflected in the
localization-delocalization transition in the group eigenmodes, as
well as, structural transformations in the
core-periphery organization of the interaction network.
We find that there is a strong correspondence between these phenomena
and the evolution of market behavior in the 90-year
interval under study, where an initial period of high stress (Great
Depression) was
followed by a long interval of relative calm but that eventually ended
in another period of high stress (Great Recession). 

Financial markets are known to exhibit periods of high volatility
leading to extreme fluctuations in asset prices which, however, seem to have little
impact on the economy as a whole~\cite{Diebold2010}.
There have been only two major exceptions to this general trend, viz.,
the 1929-33 and the 2007-09 crises, where the disturbance originating
in the financial sector spilled over to the other sectors eventually
resulting in a economy-wide slump. 
The results reported in this paper may provide significant insights
into understanding how
local perturbations propagating through complex networks can
occasionally trigger such global catastrophes.
In particular, our work suggests that loss of structural balance in
the networks representing interactions within the components of the
financial market can signal critical
accumulation of macroeconomic stress that leads to severe systemic crises. 

We would like to thank Nils Bertschinger, Bikas K Chakrabarti and 
Tiziana Di Matteo for helpful discussions. 
This work was supported in part by IMSc Econophysics (XII Plan)
Project funded by the Department of Atomic Energy, Government of
India.
ASC would like to thank Boston University where
preliminary analysis of the data was done. 
We thank the Vikram Sarabhai Library, IIM Ahmedabad for providing
access to data used in this study.

%


\begin{thebibliography}{}
\bibitem{note1}
Missing values in the price series of stocks were filled by assuming
that it was not traded
in those days such that the price remained unchanged from the last
available value.
It was ensured that stocks did not have more than 50 missing values in each interval
of length $T$ to reduce spurious correlations.
\bibitem{note3}
The eigenvector components $u_{ji}$ ($i = 1, \ldots, N$) of the $j$-th
eigenmode are normalized
such that $\sum_{i=1}^N [u_{ji}]^2 = 1$.
\bibitem{note2}
As the elements of $C^{random}$ appears to have a Gaussian distribution, imposing the specified
threshold
implies that the probability that the cross-correlation in group modes that we are using to infer
interactions could arise randomly from noise is less than 0.01.
\end{thebibliography}
\end{document}